\documentclass[pdflatex,sn-mathphys]{sn-jnl}% Math and Physical Sciences Reference Style
\usepackage{comment}
\usepackage[normalem]{ulem}
\usepackage[nolist]{acronym}
\usepackage{amsmath,bm}
\usepackage{xspace}
\usepackage{amsfonts}
\usepackage{lineno}
\usepackage[T1]{fontenc}
\usepackage[utf8]{inputenc}

\DeclareGraphicsExtensions{.png,.jpg,.pdf,.ai,.psd}
\begin{acronym}
\acro{XFEL}{X-ray Free-electron Laser}
\acro{MHz-XMPI}{MHz X-ray multi-projection imaging}
\acro{XMPI}{X-ray multi-projection imaging}
\acro{EuXFEL}{European XFEL}
\acro{SASE}{self-amplified spontaneous emission}
\acro{SPB/SFX}{Single Particles, Clusters, and Biomolecules \& Serial Femtosecond Crystallography}
\acro{DL}{Deep learning}
\acro{ONIX}{Optimized Neural Implicit X-ray imaging}
\acro{CNN}{Convolutional Neural Network}
\acro{MLP}{Multilayer Perceptron}
\acro{MSE}{Mean-square-error}
\acro{GAN}{Generative Adversarial Network}
\acro{NeRF}{Neural Radiance Field}

\end{acronym}

\definecolor{blue}{rgb}{0.07, 0.04, 0.56} 

\newcommand{\mymath}[2]{\newcommand{#1}{\TextOrMath{$#2$\xspace}{#2}}}

\mymath{\encoder}{\mathrm{\mathbf E}}
\mymath{\fcnn}{\mathrm{\mathbf F}}
\mymath{\discriminator}{\mathrm{\mathbf D}}
\mymath{\generator}{\mathrm{\mathbf G}}
\mymath{\dataDistribution}{p_D}
\mymath{\randomDistribution}{p_\nu}
\mymath{\predictContrast}{\mathbf{\hat{c}_\nu}}
\mymath{\realContrast}{\mathbf{c}_v}
\mymath{\spatialCoordinates}{\textit{\textbf{x}}}
\mymath{\refractiveIndex}{n}

\jyear{2023}%

\theoremstyle{thmstyleone}%
\theoremstyle{thmstyletwo}%

\theoremstyle{thmstylethree}%

\begin{document}

%\title[Demonstration of megahertz X-ray Tomoscopy]{Demonstration of megahertz X-ray Tomoscopy}
\title[Megahertz X-ray Multi-projection imaging]{Megahertz X-ray multi-projection imaging}

\author[1]{\fnm{Pablo} \sur{Villanueva-Perez}}\equalcont{These authors contributed equally.}\email{pablo.villanueva-perez@sljus.lu.se}

\author[2]{\fnm{Valerio} \sur{Bellucci}}
\author[1]{\fnm{Yuhe} \sur{Zhang}}
\author[2]{\fnm{\v Sarlota} \sur{Birn\v steinov\'a}}
\author[2]{\fnm{Rita} \sur{Graceffa}}
\author[2]{\fnm{Luigi} \sur{Adriano}}
\author[1]{\fnm{Eleni} Myrto \sur{Asimakopoulou}}
\author[2]{\fnm{Ilia} \sur{Petrov}}
\author[1]{\fnm{Zisheng} \sur{Yao}}
\author[3]{\fnm{Marco} \sur{Romagnoni}}
\author[3]{\fnm{Andrea} \sur{Mazzolari}}
\author[2]{\fnm{Romain} \sur{Letrun}}
\author[2]{\fnm{Chan} \sur{Kim}}
\author[2]{\fnm{Jayanath} \sur{C. P. Koliyadu}}
\author[2]{\fnm{Carsten} \sur{Deiter}}
\author[2]{\fnm{Richard} \sur{Bean}}
\author[2]{\fnm{Gabriele} \sur{Giovanetti}}
\author[2]{\fnm{Luca} \sur{Gelisio}}
\author[4]{\fnm{Tobias} \sur{Ritschel}}
\author[2,8,9]{\fnm{Adrian} \sur{Mancuso}}
\author[5,6,7]{\fnm{Henry} N. \sur{Chapman}}
\author[5]{\fnm{Alke} \sur{Meents}}
\author[2]{\fnm{Tokushi} \sur{Sato}}
\author*[5,2]{\fnm{Patrik} \sur{Vagovi\v{c}}}\equalcont{These authors contributed equally.}
%\email{patrik.vagovic@cfel.de}
%\affil[*]{Corresponding author: Darth Vader, darth@vader.deathstar}

%\authormark{12}Center for Free-Electron Laser Science CFEL, Deutsches Elektronen-Synchrotron DESY, Notkestr. 85, 22607 Hamburg, Germany\\
%\authormark{13}Department of Chemistry and Physics, La Trobe Institute for Molecular Science, La Trobe University, Melbourne, Victoria 3086, Australia\\
%\authormark{14}The Hamburg Centre for Ultrafast Imaging, Luruper Chaussee 149, 22761 Hamburg, Germany\\
%\authormark{15}Department of Physics, Universität Hamburg, Luruper Chaussee 149, 22761 Hamburg, Germany\\
%
%\authormark{7}Synchrotron Radiation Research and NanoLund, Lund University, Box 118, 221 00, Lund, Sweden\\
%\authormark{14}The Hamburg Centre for Ultrafast Imaging, Luruper Chaussee 149, 22761 Hamburg, Germany\\
%\authormark{15}Department of Physics, Universität Hamburg, Luruper Chaussee 149, 22761 Hamburg, Germany\\
\affil[1]{\orgdiv{Synchrotron Radiation Research and NanoLund}, \orgname{Lund University}, \orgaddress{\street{Box 118}, \city{Lund}, \postcode{221 00}, \country{Sweden}}}

%\address{\authormark{1}European XFEL, Holzkoppel 4, 22869 Schenefeld, Germany \\
\affil[2]{\orgname{European XFEL}, \orgaddress{\street{Holzkoppel 4}, \city{Schenefeld}, \postcode{22869}, \country{Germany}}}

%INFN Sezione di Ferrara, Via Saragat 1, 44122 Ferrara, Italy\\
\affil[3]{ \orgname{INFN Sezione di Ferrara}, \orgaddress{\street{Via Saragat 1}, \city{Ferrara}, \postcode{44122},  \country{Italy}}}

%University Collegue London\\
\affil[4]{ \orgname{University College London}, \orgaddress{\street{66-72 Gower Street}, \city{London}, \postcode{WC1E 6EA},  \country{UK}}}

%Center for Free-Electron Laser Science CFEL, Deutsches Elektronen-Synchrotron DESY, Notkestr. 85, 22607 Hamburg, Germany
\affil[5]{\orgdiv{Center for Free-Electron Laser Science CFEL}, \orgname{Deutsches Elektronen-Synchrotron DESY}, \orgaddress{\street{Notkestr. 85}, \city{Hamburg}, \postcode{22607}, \country{Germany}}}

\affil[6]{\orgname{The Hamburg Centre for Ultrafast Imaging}, \orgaddress{\street{Luruper Chaussee 149}, \city{Hamburg}, \postcode{22761}, \country{Germany}}}

\affil[7]{\orgdiv{Department of Physics}, \orgname{Universität Hamburg}, \orgaddress{\street{Luruper Chaussee 149}, \city{Hamburg}, \postcode{22761}, \country{Germany}}}

\affil[8]{\orgdiv{Department of Chemistry and Physics}, \orgname{La Trobe University}, \orgaddress{\city{Bundoora, Victoria}, \postcode{3086}, \country{Australia}}}

\affil[9]{\orgname{Diamond Light Source}, \orgaddress{\city{Harwell Science and Innovation Campus}, \postcode{OX11 0DE}, \country{United Kingdom}}}

\affil[*]{Corresponding author: \fnm{Patrik} \sur{Vagovi\v{c}}, patrik.vagovic@cfel.de}

\abstract{
%Suggestion for an abstract below 200 words:
X-ray time-resolved tomography is one of the most popular X-ray techniques to probe dynamics in three dimensions (3D).
Recent developments in time-resolved tomography opened the possibility of recording kilohertz-rate 3D movies. 
However, tomography requires rotating the sample with respect to the X-ray beam, which prevents characterization of faster structural dynamics. Here, we present megahertz (MHz) X-ray multi-projection imaging (MHz-XMPI), a technique capable of recording volumetric information at MHz rates and micrometer resolution without scanning the sample. We achieved this by harnessing the unique megahertz pulse structure and intensity of the European X-ray Free-electron Laser with a combination of novel detection and reconstruction approaches that do not require sample rotations. Our approach enables generating multiple X-ray probes that simultaneously record several angular projections for each pulse in the megahertz pulse burst. We provide a proof-of-concept demonstration of the MHz-XMPI technique's capability to probe 4D (3D+time) information on stochastic phenomena and non-reproducible processes three orders of magnitude faster than state-of-the-art time-resolved X-ray tomography, by generating 3D movies of binary droplet collisions. We anticipate that MHz-XMPI will enable in-situ and operando studies that were impossible before, either due to the lack of temporal resolution or because the systems were opaque (such as for MHz imaging based on optical microscopy).
}


\maketitle

%\section{Main}\label{sec1}
The high penetration power of X-rays makes them excellently suited to probe the make-up and dynamics of complex matter in 3D, which is  of fundamental interest for many research fields.
One of the most popular X-ray techniques to study the evolution of 3D structures is time-resolved tomography~\cite{Rolo2014Cinematography,Kareh2014Time-resolvedTomo}.
In this technique, four-dimensional (4D) - that is, three spatial plus the temporal dimension - information is acquired by gathering complete tomographic datasets at a frequency faster than the studied dynamics. These tomographic sets are for each relevant time point produced by recording several 2D images taken at different angles by rotating the sample with respect to the X-ray source (between at least $0^\circ$ and $180^\circ$).
Although time-resolved tomography can address many scientific questions \cite{Bedford:2017,Massimi:2022} and cover a wide range of spatial and temporal resolutions~\cite{Villanova2017Timeresolved,Garcia-Moreno2019NatureFoam}, it has several fundamental limitations: i) it requires rotating the sample with respect to the X-rays, thereby inducing shear forces that may alter the studied dynamics; ii) it is not compatible with single-shot or single-exposure approaches; and iii) the temporal and spatial resolution is ultimately limited by the photon flux provided by the source and the maximum scanning speed. 

The advent of \acp{XFEL} in principle opened the possibility for addressing these challenges. These facilities have  revolutionized  X-ray science and offer the opportunity to explore temporal resolutions down to the femtosecond scale and obtain high-resolution images using single pulses. Among the current \acp{XFEL}, the \ac{EuXFEL} is the first operational source with a megahertz (MHz) pulse repetition rate~\cite{Decking2020EuXFEL}.
The MHz repetition rate offers the opportunity to sample stochastic dynamics~\cite{Vagovic2019MHz} at least three orders of magnitude faster than state-of-the-art time-resolved tomography~\cite{GarciaMoreno2021kHzTomoscopy}.
However, current imaging techniques at \acp{XFEL} are only able to probe 2D information with individual pulses, unless, identical copies of the sample can be produced~\cite{Neutze2000DiffbeforeDest,Chapman2006SingleShotImaging,Ekeberg2015Virus3D}.
Only recently, a technique coined \ac{XMPI} was presented that is capable of obtaining 3D information from single pulses~\cite{villanueva2018XMPI,Duarte2019Stereo,Voegeli2020Multibeam}.
However, \ac{XMPI} has not been demonstrated before to retrieve 3D information from single \ac{XFEL} pulses.
\ac{XMPI} does away with the need for scanning the sample when acquiring volumetric information by splitting single \ac{XFEL} pulses to produce multiple beams that simultaneously illuminate the sample from different angles.

Here we report our development, implementation, and demonstration of a combination of MHz microscopy and \ac{XMPI} at \ac{EuXFEL} that, for the first time, enables studying 3D dynamics three orders of magnitude faster than state-of-the-art approaches. To that end, we implemented  \ac{MHz-XMPI} at the \ac{SPB/SFX} instrument~\cite{Mancuso:SPBSFX2019} of the \ac{EuXFEL} using:  i) a novel scheme (Fig. \ref{fig:setup}a) optimized for \acp{XFEL} beams~\cite{patent} and ii) a \ac{DL} approach to efficiently reconstruct 4D data from sparse projections. The resulting \ac{MHz-XMPI} technique is capable of probing a wide range of natural processes with micrometer  resolution and  resolving their sub-microsecond dynamics by exploiting the unique repetition rate of EuXFEL. We provide a proof-of-concept demonstration of the technique's capabilities by, for the first time, generating volumetric information of binary droplet collisions~\cite{Adam1968WaterCollision,Qian1997DropletCollData,Planchette2012DropletData,Graceffa2012DropletXRay} at MHz rates, which opens for uncovering many important natural and industrial processes, such as raindrop formation, multiphase reactors, drug delivery, ink-jet printing, and spray coating. We envision that the unique capabilities of \ac{MHz-XMPI} will enable probing many other 3D processes in materials science, medical \& clinical research, and engineering that have been obscured because of the lack of a powerful technique. 

\section{Results}\label{sec:Results}
\subsection{Overview of concept and our experimental setup}
\begin{figure}[htbp!]\centering
    \includegraphics[width=0.97\linewidth]{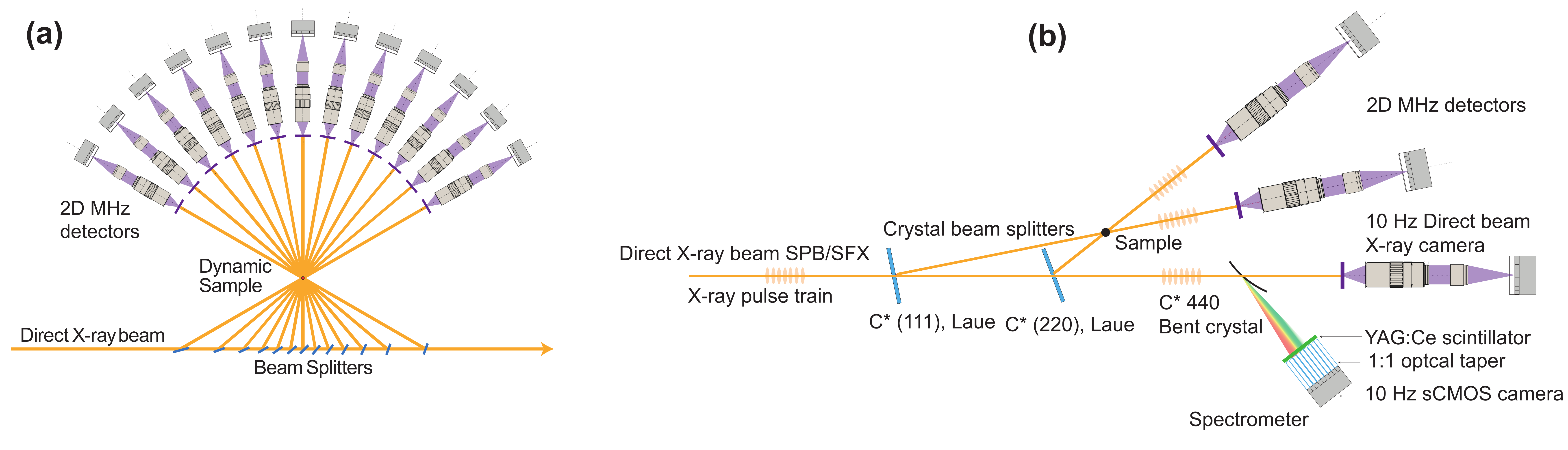}
    \caption{Conceptual scheme of X-ray Multi-Projection Microscopy for SASE beams (a), schematics of first demonstration of proposed concept with spectral and direct beam diagnostics (b).
    }
\label{fig:setup}
\end{figure}

% Introduce figure and describe the two main sets of beams
Our \ac{MHz-XMPI} concept is based on a series of beam splitters that generate different beams that illuminate from different angles, and a series of 2D MHz detectors that record the different projections [Fig.~\ref{fig:setup}(a)]. Based on this general strategy, we constructed a prototype that we used to conduct the proof-of-concept \ac{MHz-XMPI} experiments on binary droplet collisions [Fig.~\ref{fig:setup}(b)]. 

In this setup, each X-ray pulse generated by \ac{EuXFEL} is split into two beamlets in each of the crystals. One single pulse thereby splits into four beamlets [Fig. \ref{fig:setup}(b)]: i) two beamlets (also called multi-projection beamlets) that illuminate the sample from different angles and provide volumetric information; ii) a direct beam that is used for alignment and diagnostics; and iii) a beam that carries the spectral information generated by a bent crystal. 

\subsubsection{Direct and diagnostic beam}
\begin{figure}[htbp!]\centering
    \includegraphics[width=0.97\linewidth]{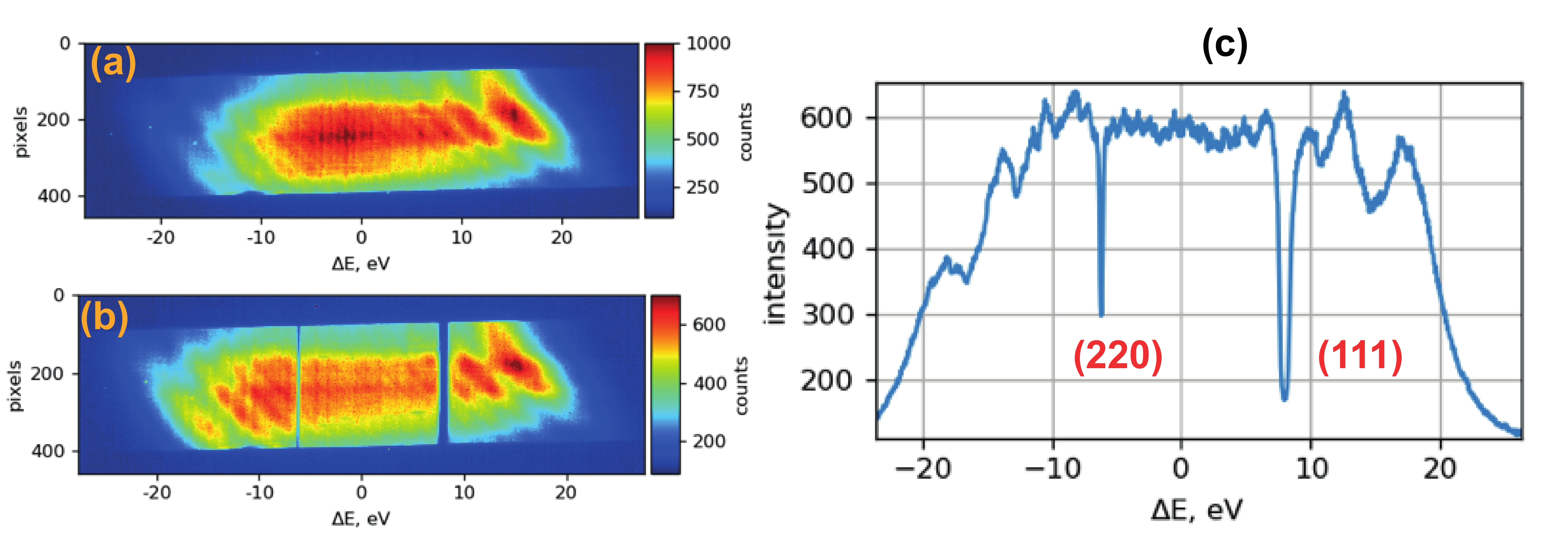}
    \caption{The spectrum diagnostics. 2D image of the SASE beam without beam splitters (a) and with two Diamond (111) and (220) crystals in the beam (b). The horizontal line profile with two spectral dips of missing intensity are clearly visible (c).}
\label{fig:spectrum}
\end{figure}

The direct or diagnostic beam is used for two primary purposes: spectral and spatial alignment of the crystals. In our setup Fig.~\ref{fig:setup}(b), the direct beam passes through an energy-dispersive element consisting of a bent crystal, which diffracts different portions of the direct beam into different directions depending on their energy component. The photon energy components of each diffracted beam are monitored by a 10 Hz sCMOS camera with dispersion axis aligned along the horizontal direction of the detector. After the photon energy calibration ~\cite{Petrov:2023}, each pixel along the diffraction plane corresponds to a specific photon energy with a given bandwidth or energy resolution. In turn, the spectrometer monitors the portion of the beam that is diffracted by each of the multi-projection beamlets. Using this approach enabled us to optimize the efficiency and imaging performance of the \ac{MHz-XMPI} setup in our experiments by selecting the highest spikes of the \ac{SASE} spectrum over the lower-amplitude background. Moreover, the spectrometer (see Methods \ref{sec:spectrometer}) enabled us to evaluate the performance of the crystal splitters by providing a direct measurement of the extracted bandwidth and efficiency from the direct beam. The resulting spectrum profile from our characterization revealed the spectral dips corresponding to the two crystal splitters that generated the multi-projection beamlets in our experiments Fig.~\ref{fig:spectrum}. From this spectrum profile, we could estimate and optimize the imaging capabilities of the multi-projection beamlets, i.e.,
maximize the fluence of each beamlet by positioning them along the most intense components of the SASE spectrum and preventing their spectral overlap.

To align all of the optical components in the direct beam and spatially align the crystal splitters in our experiments, we projected the direct beam transmitted through the bent crystal onto an indirect-imaging detector operating at 10 Hz, the train frequency of the \ac{EuXFEL}. To achieve further fine-tuning of the angular and spatial alignment of the crystal beam splitters, we used the spectrometer to select the high-intensity component of the \ac{SASE} spectrum and maximize the diffracted intensity.

\subsubsection{Crystal splitters and split beams} 
To generate the multi-projection beamlets, we used two synthetic-diamond crystals (see Methods \ref{sec:splitters}) that were chosen because of i) their resilience to the high-intensity pulses of \acp{XFEL}, ii) high transparency because of the low atomic number of diamond (C), and iii) high purity, near-lack of dislocations, and low strain~\cite{Tarelkin:2016}. For our experiments, we selected as beam splitters two diamonds with reflections (111) and (220) in the Laue configuration [see Fig. (\ref{fig:setup}b)]. Because the \ac{EuXFEL} was operated at 10~keV, the C(111) and C(220) deflected the beam $35.0^{\circ}$ and $58.8^\circ$, respectively, with respect to the direct beam. The resulting relative angle between the two split beams of $23.8^{\circ}$ was used to obtain volumetric information. The alignment of the crystal splitters to ensure that the two beams intersected in the sample position to obtain volumetric information from a single pulse and the optimization of the diffracted flux (spatial and spectral optimization) were done simultaneously in an iterative manner. Although we only generated two multi-projection beamlets in the experiment, the setup can potentially generate more beamlets from a single \ac{XFEL} pulse by using more spectrally-optimized crystal splitters that exploit other high-intensity peaks of the \ac{SASE} spectrum, see Fig~\ref{fig:setup}(a). 

\subsection{Proof-of-concept demonstration on binary droplet collisions}
% Sample and detection
To provide a first proof-of-concept demonstration of the strengths and capabilities of our \ac{MHz-XMPI} technique, we used the experimental setup described above to study binary droplet collisions~\cite{Adam1968WaterCollision,Qian1997DropletCollData,Planchette2012DropletData}. This is a system of great importance for natural and industrial processes, such as raindrop formation~\cite{Zaillang1995RainDrop}, multiphase reactors~\cite{Faeth1977MultiPhaseReactors}, drug delivery~\cite{Planchette2010DrugDelivery}, ink-jet printing and spray coating~\cite{ashgriz1990Spray,Qian1997DropletCollData}. However, despite the importance of these processes, they have so far only been studied in 2D with MHz rates and mainly with visible light, which does not capture the volumetric information of the rich dynamics contained in these processes. Specifically, the droplets are not fully transparent to visible light, which prevents gathering all relevant information that is necessary to understand these processes. Our proof-of-concept demonstration of the \ac{MHz-XMPI} technique is thus the first experiment that provides volumetric information at MHz rate with X-rays (a radiation that can penetrate the droplets) on binary droplet collisions.

\subsubsection{Spatial and temporal aligning}
To enable retrieving MHz volumetric information, we aligned the binary droplet setup (see Methods \ref{sec: ballistic droplets}) to the point of intersection between the two beamlets to ensure that the binary droplet collisions were illuminated by both beamlets. This required not only a spatial, but also temporal, alignment of the droplet injection to the arrival of the X-ray pulses from the \ac{EuXFEL}. The temporally and spatially aligned images provided by each split beamlet were recorded by a pair of indirect detectors. The indirect detection system (see Methods \ref{sec: detectors}), after being synchronized to the arrival of each split beamlet, enabled 1.128 MHz acquisitions with $3.2~\mu$m and $6.4~\mu$m pixel sizes.

\subsubsection{Image processing and reconstruction using deep learning}
\begin{figure}[htbp!]\centering
    \includegraphics[width=0.97\linewidth]{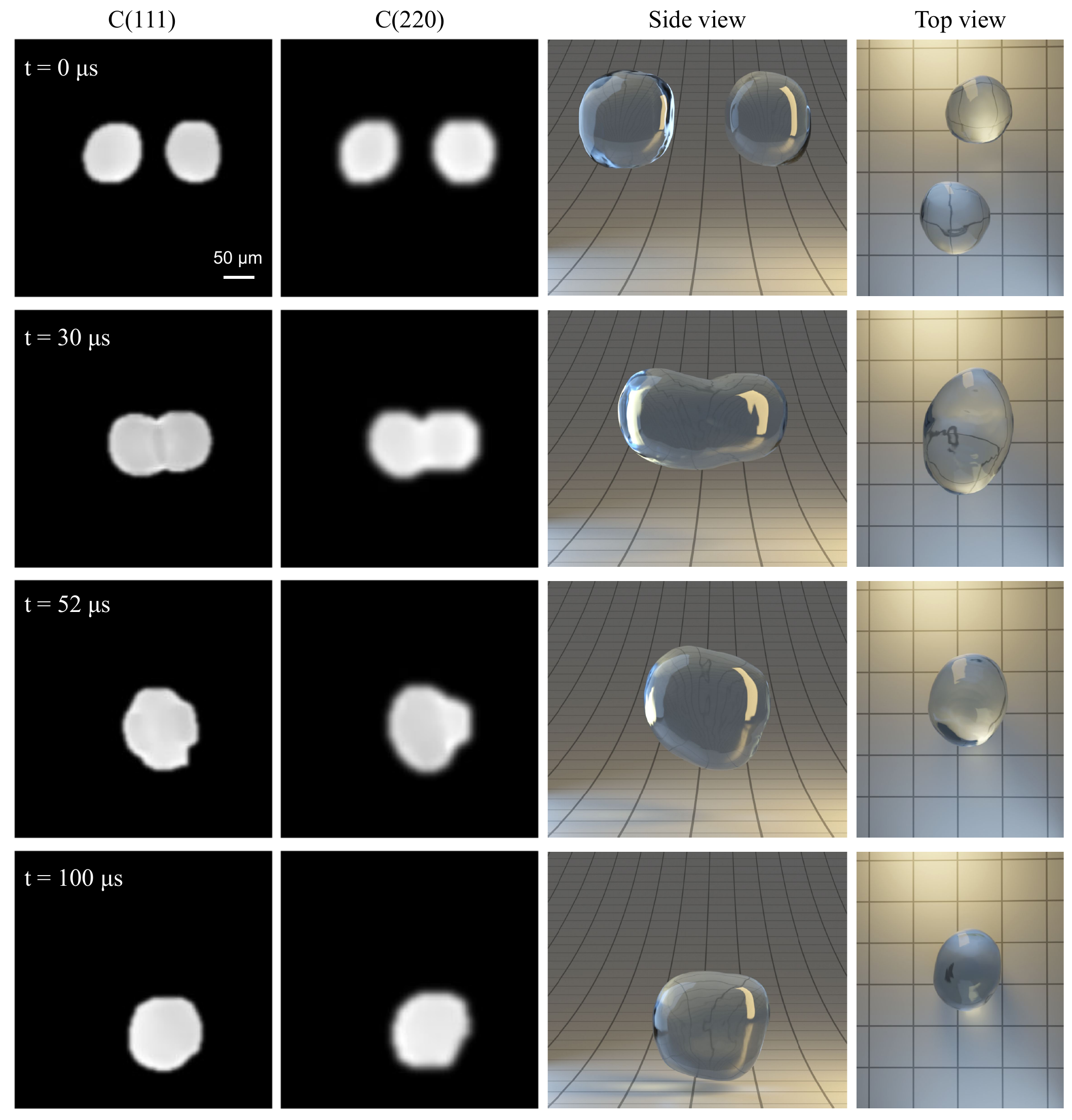}
    \caption{Binary droplet collision. The first two panels depict the projections from C(111) and C(220) for different relevant time points of the collision used to reconstruct 3D movies. The right panels show the rendering of the 3D movie corresponding to the selected time points. Two views are shown. The side view is on the plane of the collected views, while the top view is perpendicular to both of the acquired projections.}
\label{fig:data}
\end{figure}

% Introduce the image processing and the reconstruction
To enable producing 3D movies of the dynamic processes recorded using the setup described above, we first processed the recorded projections by correcting the shot-to-shot noise arising from the stochastic nature of the \ac{SASE} process.  This correction was performed using principal component analysis~\cite{VanNieuwenhove15SVDFFC,Buakor22FFC} (see Methods \ref{sec:processing}).
The corrected images were then filtered using wavelet decomposition and total-variation regularization to reduce the noise component.
% Flat field correction and filtering
From those filtered and corrected images, the droplets were segmented from the background and used to retrieve 4D reconstructions.
From the two processed projections for four different time points (left panels in Fig.~\ref{fig:data}) we found that the images from the C(220) beamlet were noisier and have worse contrast than the ones of the C(111). We attribute this to the reduced diffraction efficiency because of using a higher diffraction order and lower reflectivity due to polarization effects (see Discussion).

% 4D ONIX
Using the processed images, we reconstructed a 3D movie by combining the two projections into a 3D volume (see Methods \ref{sec:reco}) and using a self-supervised \ac{DL} algorithm based on \ac{NeRF}~\cite{Mildenhall2020NERF} and our approach \ac{ONIX}~\cite{Yuhe2022ONIX}, which can reconstruct 3D information from sparse projections and temporal information. We achieved this cumbersome task by: i) including a physical model of the X-ray interaction with matter for image formation, ii) having a 4D functional formalism to describe the sample as a function of position and time, iii) transferring knowledge across different experiments and processes, and iv) enforcing consistency between the 4D functional description and the recorded projections by the two beamlets (more details about our algorithm can be found in the Supplementary Information).

We used two independent binary droplet collisions, each of them with up to 127 frames per projection or beamlet, to train the algorithm and produce the results presented in Fig.~\ref{fig:data}. The whole rendered movie for one of those processes at 1.128~MHz and a voxel size of $3.2~\mu$m can be found in the Supplementary Information. These results represent the first 3D movie acquired using single shots of an \ac{XFEL} with a temporal resolution that is three orders of magnitude faster than state-of-the-art time-resolved tomography.

\section{Discussion}\label{sec:discussion}
Our proof-of-concept experiment is the first demonstration of the \ac{MHz-XMPI}, which exploit the unique repetition rate and high-intensity pulses of the \ac{EuXFEL} to record MHz processes with micrometer resolution, three orders of magnitude faster than state-of-the-art 4D X-ray imaging techniques. We achieved this using a novel splitter configuration (Fig. \ref{fig:setup}(a)) and reconstruction algorithm at the \ac{SPB/SFX} instrument of the \ac{EuXFEL}. 

To demonstrate the capabilities of the technique, we imaged binary droplet collisions using a photon energy of 10~keV at the \ac{EuXFEL}. We recorded two independent projections of the collisions, separated by $23.8^{\circ}$, with 1.128~MHz temporal sampling and pixel sizes of $3.2~\mu$m and $6.4~\mu$m using single pulses of the \ac{EuXFEL}. Although the recorded repetition rate was 1.128~MHz, one can reach up to 4.514 MHz using the maximum pulse repetition rate of the \ac{EuXFEL}, without altering the image quality. In turn, the pixel size can potentially be further reduced to $1.6~\mu$m in the current detector. However, the achievable spatial resolution is not limited by the pixel size, but rather by the number of photons per area per pulse, the efficiency of the beam splitters, and the sample contrast. Thus, the resolution has to be estimated on a sample-to-sample basis, and it can be improved by reducing the detector pixel size and increasing the fluence per pulse by, e.g., focusing the \ac{XFEL} beam. For these presented experiments, the beam was not focused and maintained its natural divergence of few $\mu$rad~\cite{Mancuso:SPBSFX2019}.

The resolution and contrast we achieved in our proof-of-concept demonstration already enabled filming binary droplet collisions because the droplets have sizes around 70-80~$\mu$m. The relevant parameters to characterize such collisions are the Weber number~\cite{Sirignano:Droplet1999} and impact parameter (the ratio between center-to-center distance measured perpendicularly to the relative velocity direction averaged and the averaged droplet diameter). For the studied collisions, we estimated a Weber number of around 7.4 and an impact parameter of 0.12 (almost head-on collisions). The relative velocity of the two droplets was around 2.4 m/s, which corresponded to a displacement of less than a pixel between consecutive frames at 1.128 MHz. This demonstrates that \ac{MHz-XMPI} can capture in 3D the dynamics of this process at relevant spatial and temporal resolution. Because of the penetration power of X-rays we could thereby obtain real volumetric information that was previously impossible to capture in most of the previous experiments using visible light~\cite{Rita_Thesis:2012}. Taken together, our demonstration is the first real-volumetric investigation of binary droplet collisions with X-rays at relevant spatial and temporal resolutions.

% Third paragraph challenges and work to do
%5) Challenges: 
% Instrumentation:
%synchronization and optimization of the splitting setup
%6) Acquisition with MHz system.
% Software:
%7) Reconstruction from the sparse projection
Achieving such unique temporal and spatial resolutions with \ac{MHz-XMPI} required us to address several instrumentation and reconstruction challenges. First, for instrumentation, the most crucial challenge was to achieve efficient splitting of single pulses generated by an \ac{XFEL}. We achieved this using diamond crystals and optimizing them using a spectrometer (see Methods \ref{sec:splitters} and \ref{sec:spectrometer}). To address the issue of the Laue-configuration-oriented diamond splitters displacing the beam in the detector plane around a few tens of micrometers from shot-to-shot as a result of the fluctuations of the \ac{SASE} spectrum, we registered the recorded movies for each detector independently (see Methods \ref{sec: detectors}).

Second, we needed to address challenges related to synchronization and reconstruction of the images. Specifically, each detector was synchronized to the arrival of the beamlets and the experiment itself to the arrival of the \ac{XFEL} pulses (see Methods \ref{sec: detectors}). The main reconstruction challenge was how to retrieve volumetric information from only a few projections (in our case, we measured two projections), where conventional time-resolved tomography typically relies on hundreds or thousands of recorded projections for every time point equally spaced between $0^{\circ}$-$180^{\circ}$. We addresses this challenge through a \ac{DL} approach that includes the physics of image formation and transfers information across experiments (see Methods \ref{sec:reco}).

% Open challenges
Our first proof-of-concept demonstration of \ac{MHz-XMPI} can be further built on and improved  to fully exploit the capabilities of this technique. To do so, we will address several challenges in future iterations of our current experimental setup.
% Air path loses
First, it will be important to decrease the distance the beam has to travel through air in the experimental setup. In our case, the experimental setup was not compact, with approximately 2 m of air that the beam had to pass, which caused around 70~\% loss of the photon fluence per pulse when using 10~keV X-ray photons.
% Detectors
Second, light collection could be increased by using lenses in the indirect detection system with higher numerical apertures. This could lead to higher resolutions and contrast. Nonetheless, the contrast (phase contrast, absorption contrast) is also limited on the camera side by its low dynamic range of 10 bits. Increasing the dynamical range from actual 10 bits to 16 bits while reducing pixel size or increasing the format of the chip would provide a significant increase in  sensitivity and spatial resolution for multi-projection phase contrast imaging at MHz rates. This would require the development of a new generation of MHz cameras.  
% Image doubling
Third, the distance between the high-intensity peaks of the \ac{SASE} spectrum is also a challenge that needs further developments. For a few shots, two or three peaks are simultaneously accepted by a crystal splitter, which causes a doubling or tripling of the image in the Laue configuration. This could be corrected by using crystals in symmetric Bragg~\cite{Bushuev:BraggLaueXFEL2008} or using self-seeding to increase the spectral density and lock the energy~\cite{Feldhaus:Selfseeding1997,Amann:SelfseedingDemo2012}. 
% Increasing projections and efficiency.
Fourth, we aim to develop new splitting configurations that increase the number of projections acquired, their angular separation, and the fluence per split beamlet. For such a purpose, we will explore new splitting schemes, new crystal materials, and configurations.
%Sigma vs pi 
Finally, the current setup uses crystal splitters with diffraction planes oriented in the horizontal plane and X-ray electric field polarised in the same plane to form a so-called $\pi$--configuration. Such arrangement has 10\% - 20\% narrower rocking curves as compared to its perpendicular counterpart ($\sigma$--configuration)~\cite{Authier2001}. Moreover, the $\pi$--configuration is modulated by polarization factor $P = \cos^2(2\theta)$, where $\theta$ denotes the Bragg angle,  while in case of $\sigma$-polarisation $P = 1$. Hindering the use of the $\pi$--configuration to generate large splitting angles compared to the $\sigma$--configuration. As such, one could consider using the $\sigma$--configuration to improve the throughput of the system, especially for large angular separation. We will incorporate all of the above-mentioned optimizations in future iterations of our experimental setup, to maximize the throughput of the system and the number of projections to generate significant improvements in image quality and enable access to higher photon energies that are important to image samples with higher densities.  

% Fourth paragraph
%9) Conclusions and outlook
Taken together, we have demonstrated the capabilities of the \ac{MHz-XMPI} to exploit the unique repetition rate and high-intensity pulses of the \ac{EuXFEL} to record MHz processes with micrometer resolution, three orders of magnitude faster than state-of-the-art 4D X-ray imaging techniques. 
% Outlook
The \ac{MHz-XMPI} technique has thereby the potential to revolutionize time-resolved imaging in opaque samples by providing 3D movies with micrometer and sub-microsecond resolutions (detectable object velocities up to km/s range), which have not been possible before until now with X-ray imaging. Examples of fields of research and technology that benefit from \ac{MHz-XMPI} are additive manufacturing, bio-printing, new material production, fundamental high-speed dynamics in fluidics, bubble formation and collapse in propulsors, flow mixing in chemical engineering and fuel cells, combustion within engines, ultrasound therapy in biomedicine, observing a beating heart in real time, crack propagation through a material under stress conditions, penetration of the drugs into the body, as well as shockwaves induced by ultrasound treatment and their real-time influence on the tissues and many more.

\section{Methods}\label{sec:methods}

\subsection{Crystal beam splitters}\label{sec:splitters}
The diamond high-pressure and high-temperature (HPHT) dislocation-free crystals with a thickness of 100 \textmu m have been used as beam splitters.
Each crystal was aligned along the direct beam using high-precision 6-degree-of-freedom SmarAct manipulators. The first splitter with the surface orientation (110) was set in symmetric Laue configuration using the (111) reflection aligned in the horizontal diffraction plane. The second crystal was (100) oriented, and the symmetric Laue (220) reflection was aligned at the same horizontal diffraction plane as the previous splitter. The diffracted beams from each crystal have been aligned to overlap at a common interaction point where the sample was placed.  

\subsection{Spectrometer}\label{sec:spectrometer}
Downstream the crystal splitters, a bent diamond crystal with (110) oriented surface was aligned to diffract in the horizontal plane using the (440) reflection in Bragg configuration. The bent crystal acts as a spectrum analyzer \cite{Makita:15,Boesenberg:17}, and in this experiment, its dispersion axis was oriented in the horizontal plane. An Andor Zyla 5.5 HF sCMOS camera with optical 1:1 taper with a 50 \textmu m thick YAG:Ce scintillator was used to record the spectrum at 10 Hz, i.e., the train repetition rate of \ac{EuXFEL}. 
The photon energy and resolution were determined using the absolute wavelength calibration method \cite{Petrov:2023}. 
When a splitter is inserted into the beam and set to diffraction condition, a dip appears in the spectrum, which is used for the precise positioning of each splitter in the spectral space of SASE pulses (Fig. \ref{fig:spectrum}).

\subsection{Binary droplet collisions}\label{sec: ballistic droplets}
The two water droplet streams are ejected from two piezo-driven microdrop generators (Microdrop Technologies GmbH) with a frequency of 10 Hz, in phase with the \ac{EuXFEL} trains. Constant pressure is kept on the sample reservoirs and the fluid lines through a Microfluidic Flow Controller (Elveflow) to ensure stable operations. Furthermore, the setup allows for refilling the reservoirs and flushing the microdrop heads remotely.
A fast light-emitting diode (HARDsoft), also triggered in phase with the XFEL trains, provides stroboscopic illumination for optical microscopy observation. It ensures the correct sample placement in the field of view and synchronization with XFEL trains. Moreover, the two droplet stream ejections are triggered independently, allowing to set a relative delay among them to merge the droplets in flight. The merging of the two droplets is furthermore facilitated by additional translation motors equipping one microdrop head.~\cite{Graceffa2012DropletXRay}

\subsection{Detection}\label{sec: detectors}
To record X-ray images for individual X-ray pulses for each beamlet and the transmitted direct beam (see Fig. ~\ref{fig:setup}(b)), three indirect detectors have been used. Each detector consists of a scintillator, $45^{\circ}$ mirror, infinity corrected objective, and a motorized insertable optical prism splitter, which creates two optical branches and enables the connection of two independent cameras to the microscope. Each branch contains two tube lenses translating the latent image from the scintillator into the corresponding sensor plane of each camera. 
The detectors for the multi-projection beamlets used a MHz-rate Shimadzu Hyper Vision HPV-X2 FT-CMOS camera that was connected to one optical branch and a Basler camera to the other optical branch for alignment purposes. For the direct beam detector, an Andor Zyla 5.5 camera was used for the spatial alignment of the splitters and the bent crystal. The microscope imaging the beamlet generated by  C(111) and the direct beam used a Mytutoyo $10\times$ NUV 0.28 NA objective, while the microscope imaging the C(220) beamlet used a Mytutoyo $5\times$ 0.21 NA HR objective to compensate for the lower flux from this reflection. In all microscopes, Mytutoyo MT-L4 tube lenses were used in each optical branch. 
MHz-rate cameras have been synchronized to the same pulse train containing 128 X-ray pulses, with each image in both camera buffers aligned to emission light from the scintillators. 
The synchronization of the imaging system was done in the following way. HPV-X2 camera is synchronized by 10 Hz TTL signals, standby, and trigger. 
These signals are provided by the \ac{EuXFEL} timing system based on MicroTCA technology issued before the train's arrival. 
%Cameras outputs configurable TTL signal, which was set to provide logical high for each frame while exposing and low in between exposures. 
To temporally locate the X-ray pulses, we used a Silicon diode, which received X-ray scattering signal from each pulse. 
The signals have been connected to the oscilloscope, and the standby and trigger signals have been delayed; as such, the first X-ray pulse in a train was aligned to the second frame of the camera. 
The reason for this is that the first frame of the camera contains strong electronic noise reaching 50\% of the dynamic range. 
Therefore only 127 of the 128 pulses were used. 
%Due to the clock mismatch of the camera and EuXFEL timing system, the full synchronization at this state is not possible. 
1.128 MHz pulse repetition rate (886 ns pulse separation) of the \ac{EuXFEL} was chosen as the camera repetition rate (890~ns) is very close to this frequency.
%the \ac{EuXFEL} pulse frequency (886~ns) with frame separation of 890 ns while X-ray pulse separation is 886ns. 
This mismatch caused a 4~ns relative drift of the camera frame to X-ray pulse for every successive frame, leading to a cumulative drift of 508~ns, which was enough to keep all 127 frames within the exposure window of the camera. 
%The maximum available exposure time at the frame spacing of 890ns was 590ns which was enough to keep all 128 frames within the exposure windows of each camera frame.

\subsection{Image processing}\label{sec:processing}
The two projections recorded by the detection system were processed before being used for the 4D reconstructions.
First, we performed flat-field correction using PCA component analysis~\cite{VanNieuwenhove15SVDFFC,Buakor22FFC}.
Seven principal components were used to flat-field correct the projections together with a downsampling operation with factors $(2,4)$ to accelerate the computation.
The flat-field corrected images for both detectors were then filtered.
First, we performed up to three shifted cycles (cycle\_spin) of wavelet shift-invariant filtering with db1 wavelets and sigma re-scaled.
Then, we used total-variation denoising using split-Bregman optimization with a denoising weight of 0.9 and a maximum of 100 iterations with a stopping criterion of 0.001.
The implementations for the wavelet and total-variation filtering were from the Python package Scikit-image~\cite{van2014scikit}. 
The denoised images of the beamlet from the C(111) splitter were then registered to  compensate for the shot-to-shot jitter using mutual information and affine transformations.
The second beamlet did not require registering as the jitter was not obvious in the images because of the resolution and spectral width of the splitter.
The second beamlet image was resized to the pixel size of the C(111) beamlet, as the pixel size of the latter was half the size of the former.
The resulting images were then cropped to the region of interest where the collisions took place.
The size of the cropped patches was 128$\times$128.
Finally, a mask to separate the background from the droplets was determined by using a high-pass filter and a canny operator to detect the edges.
Some results of this image-processing pipeline are depicted in Fig.~\ref{fig:data}.

\subsection{4D reconstructions and rendering}\label{sec:reco}
We reconstructed 3D movies from the processed two projections, using a deep-learning multi-projection reconstruction method based on neural implicit representations~\cite{Mildenhall2020NERF}, which expresses the complex refractive index of the droplets as a function of the 4D spatial-temporal coordinates. 
We exploited \ac{ONIX}~\cite{Yuhe2022ONIX} networks to learn the implicit function.
\ac{ONIX} consists of two networks: i) an encoder network that generalizes over different instances to extract common latent features from the projections and ii) a fully-connect neural network that maps the 4D coordinates and latent features to the refractive index.
The outputs of the second network (refractive index value at each point) were integrated along each line of X-ray propagation, providing predictions  at each query angle. 
The networks were trained in an unsupervised way using \ac{GAN}, where a third network was added as a discriminator to learn the distribution between the real processed projections and the network predictions. 
Refer to supplementary document 1 for details of the reconstruction algorithm.
After the training, we obtained volumetric information on the collision of the droplets by extracting 512$^3$ grids from the learned implicit function for each XFEL pulse.
Those grids were filtered using a Gaussian kernel with a sigma equal to 2 pixels and downsampled to a 128$^3$ grid that matched the recorded pixel resolution.
The resulting grids were rendered using Blender (see Fig. ~\ref{fig:data}).

%%%%%%%%%%%%%%%%%%%%%%%%%%%%%%%%%%%%%%%%%%%%%%%%%%%%%%%%%%%%%%%%
% Back matter with the required information by Nature Photonics
%%%%%%%%%%%%%%%%%%%%%%%%%%%%%%%%%%%%%%%%%%%%%%%%%%%%%%%%%%%%%%%%
\backmatter

\bmhead{Supplementary information}

Supplementary discussion: 4D MHz-XMPI reconstruction algorithm from sparse projections.

%\section*{Declarations}

\section*{Data availability}
The experimental data that support the current study are available in the following public repository \href{https://zenodo.org/}{Zenodo to be added in editorial step}.

\section*{Code availability}
%standard sentence
All codes used to produce the findings of this study are available on GitHub. The codes to process the raw data and generate the 4D reconstructions can be found in~\href{https://github.com/pvilla/MHz-XMPI_dataprocessing.git}{GitHub MHz data processing repository} and~\href{https://github.com/yuhez/ONIX_XMPI_Recon.git}{GitHub ONIX XMPI repository}, respectively.

\section*{Acknowledgments}

%Acknowledge EuXFEL for the beamtime
We acknowledge European XFEL in Schenefeld, Germany, for the provision of X-ray free-electron laser beamtime at Scientific Instrument SPB/SFX and thank the staff for their assistance.
This work received funding by HORIZON-EIC-2021-PATHFINDEROPEN-01-01, MHz--TOMOSCOPY project, Grant agreement: 101046448;
ERC-2020-STG, 3DX-FLASH, Grant agreement 948426; EuXFEL R\&D “MHz X-ray microscopy: From demonstration to method”, 2020 - 2022;Bundesministerium für Bildung und Forschung (BMBF) (05K18XXA),Vetenskapsrådet (VR) (2017-06719), Röntgen Ångström Cluster INVISION project. PV acknowledge Thomas Dietze for technical support. 

\section*{Author contribution}
Measurements (PVP, VB, YZ, SB, RG, LA, EMA, IP, AM, RL, CK, JK, RB,TS and  PV),  instrumentation and infrastructure development (VB,SB, RG,LA,EMA,IP,MR,AM,RL,CK,JK,CD,RB,GG,LG,APM, AM,TS and PV), experimental design (PVP, VB, EMA and PV), binary droplet collision setup (RG and LA), analysis (PVP, YZ, SB, IP, ZY, TR), concept (PVP, PV). PV and PVP wrote the manuscript with input from all other authors. 

\section*{Competing interests}
PV, VB, PVP, and Wataru Yashiro, Tohoku University,  filed a patent based on some key aspects described in the article (application no. EP21200564.9, 1. 10. 2021). Other authors declare no competing interests. 
%Some journals require declarations to be submitted in a standardised format. Please check the Instructions for Authors of the journal to which you are submitting to see if you need to complete this section. If yes, your manuscript must contain the following sections under the heading `Declarations':

%\begin{itemize}
%\item Conflict of interest/Competing interests (check journal-specific guidelines for which heading to use)
%\item Ethics approval 
%\item Consent for publication
%\item Availability of data and materials
%\item Authors' contributions
%\end{itemize}

%\noindent
%If any of the sections are not relevant to your manuscript, please include the heading and write `Not applicable' for that section. 

%\bigskip

%\bigskip\noindent
%Springer journals and proceedings: \url{https://www.springer.com/gp/editorial-policies}

%\bigskip\noindent
%Nature Portfolio journals: \url{https://www.nature.com/nature-research/editorial-policies}

%\bigskip\noindent
%\textit{Scientific Reports}: \url{https://www.nature.com/srep/journal-policies/editorial-policies}

%\bigskip\noindent
%BMC journals: \url{https://www.biomedcentral.com/getpublished/editorial-policies}
%\end{flushleft}

\begin{appendices}

\section{4D MHz-XMPI reconstruction algorithm}\label{secA1}

\begin{figure}[htbp!]\centering
    \includegraphics[width=0.97\linewidth]{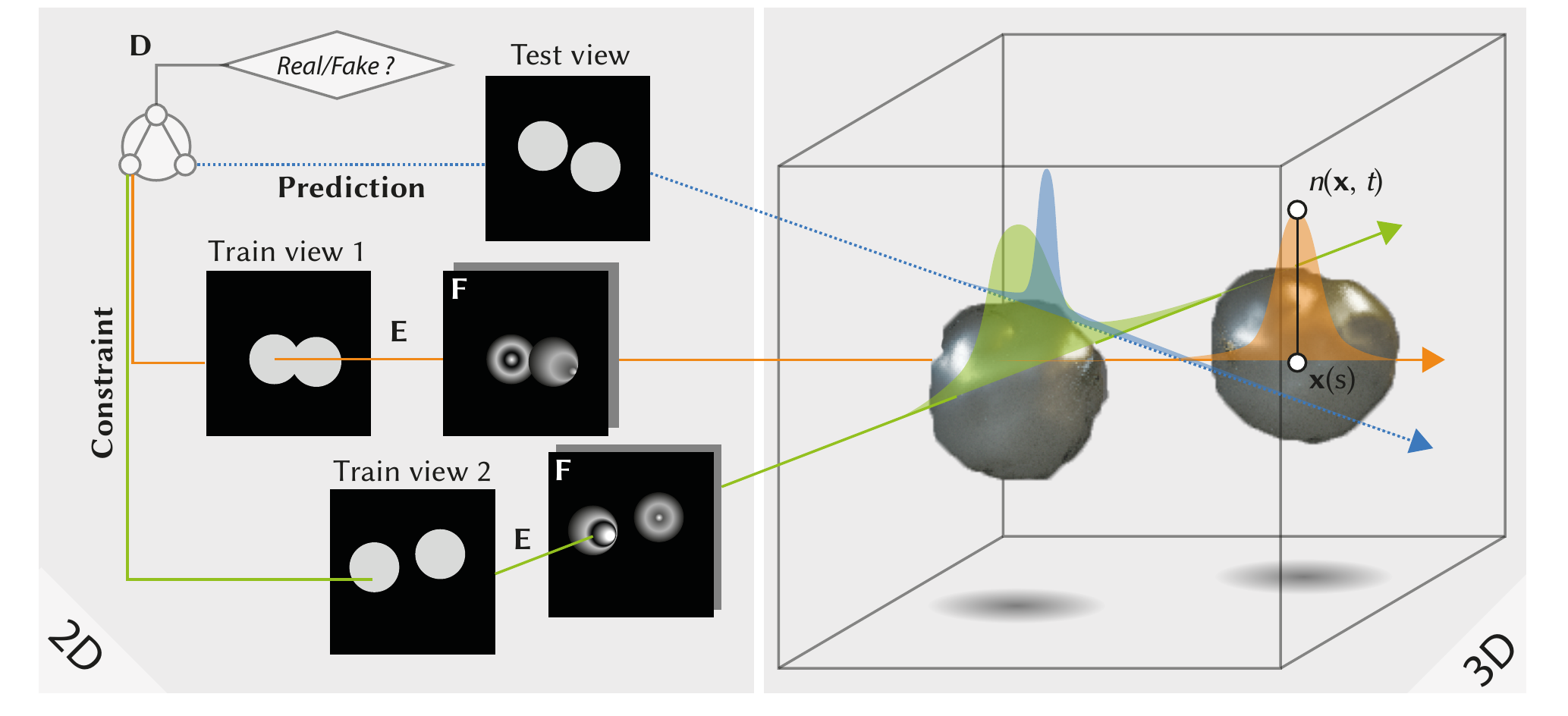}
    \caption{Overview of the reconstruction approach.}
\label{fig:4DONIX}
\end{figure}

In this appendix, we briefly describe the deep-learning algorithm used for reconstructing 3D movies of the droplets from the sparse views recorded by \ac{MHz-XMPI}.
The movies were reconstructed using an approach based on \ac{ONIX}~\cite{Yuhe2022ONIX}, a 3D multi-projection reconstruction approach that applied implicit neural representation~\cite{Mildenhall2020NERF,yu2021pixelnerf} to learn the mapping between the spatial coordinates \spatialCoordinates and the complex refractive index \refractiveIndex.
In this work, as a time sequence was acquired, the goal was to learn the mapping from 4D spatial-temporal coordinates to the refractive index: $\left ( \spatialCoordinates, t \right ) \mapsto \refractiveIndex $, as illustrated in Fig. \ref{fig:4DONIX}.
\ac{ONIX} contains two key components.
The first component is an encoder (\encoder) implemented by a \ac{CNN}. 
\encoder encodes the recorded projection images (constraints) into latent space, thus, learning common features of the data and generalizing over different instances or experiments.
The second component is a fully-connected neural network (\fcnn). 
\fcnn uses as input the encoded features from \encoder and the 4D coordinates in the reconstruction frame to provide a complex index of refraction as a function of space and time.
In the \ac{MHz-XMPI} experiment, we only collected absorption-contrast images.
Therefore, we reconstructed only the absorption component of the index of refraction and not the phase component.
Predictions at any query angle can be calculated by integrating the output along the line that corresponds the propagation direction of an X-ray ray, following the law of projection approximation~\cite{Paganin2006CoherentX-ray}.
Refer to \cite{Yuhe2022ONIX} for more details of \ac{ONIX} architecture.
In this experiment, we wanted to use both of the two projections as constraints to make full use of the data we collected, so we trained \ac{ONIX} in an adversarial way~\cite{schwarz2020graf, henzler2019escaping}, using \ac{GAN}~\cite{goodfellow2020generative, mirza2014conditional}.
For such a purpose, a discriminator (\discriminator) was included.
\discriminator learns to minimize the difference between the distribution of the real projections and the predictions.
For each time point, we computed predictions from all view angles in the detector plane.
The discriminator \discriminator saw image patches from both the real projections and the fake ones (\ac{ONIX} output), and it was trained to distinguish the fake ones from the real data.
The encoder and the fully-connected neural network received feedback from the discriminator and learned to fool the discriminator by generating predictions that are difficult to distinguish.

The loss function is expressed in Eq.~\ref{eq:loss}.

\begin{equation}
{\mathcal L_\mathrm{GAN}} = \mathbb{E}_{\realContrast \sim \dataDistribution}\log(\discriminator(\realContrast))+\mathbb{E}_{\predictContrast\sim \randomDistribution}\log(1-\discriminator (\predictContrast))~,
\label{eq:loss}
\end{equation}
where \realContrast and \predictContrast denote image patches from the real and predicted projections, respectively.
\dataDistribution denotes the distribution over the collected projections in our experiments, which are considered real projections by the discriminator.
While \randomDistribution denotes the distribution over all generated predictions, which are considered fake projections by the discriminator. 
$\mathbb E$ denotes the expectation.
We used the same sampling method as in \cite{schwarz2020graf} to extract image patches, where each image patch was formed by sampling a $32\times32$ square grid with a flexible scale, position, and stride.

We used ONIX generators~\cite{Yuhe2022ONIX} and PatchGAN discriminator~\cite{ledig2017photo} for the training.
The adversarial loss was applied after five epochs to avoid the mode collapse, a well-known problem of \acp{GAN}~\cite{kodali2017convergence}.
In the first five epochs, the network only learned the self-consistency over the two given views. 
This was enforced by minimizing the \ac{MSE} between the prediction and the input of the given views.
We also applied positional encoding~\cite{Yuhe2022ONIX,Mildenhall2020NERF,vaswani2017attention} to \spatialCoordinates and $t$ to map from 
low-dimension spaces $\mathbb{R}$ to high-dimension spaces $\mathbb{R}^{2L}$.
Specifically, we used $ L_\spatialCoordinates =10 $ and $L_t = 6$ for the spatial and time coordinates, respectively.
The Adam optimizer~\cite{kingma2014adam} was used with a batch size of six.
The learning rates were set to 0.0001 for all networks. 
We performed the training on the European \ac{XFEL} computing server using a single NVIDIA A100 GPU with 40~GB of RAM. 
The presented reconstructions were the results after two hours of training.




\end{appendices}


\bibliography{master}% common bib file


\end{document}